\documentclass[prb,twocolumn,showpacs,floatfix]{revtex4}
\usepackage{graphicx}
\usepackage{dcolumn}
\usepackage{bm}
\begin{document}
\title{Magnetoelastic effects in multiferroic YMnO$_3$}
\author{Tapan Chatterji$^1$,   Bachir Ouladdiaf$^1$, Paul F. Henry$^2$, and Dipten Bhattacharya$^3$}
\address{$^1$Institut Laue-Langevin, B.P. 156, 38042 Grenoble Cedex 9, France\\
$^2$European Spallation Source, ESS AB, P.O. Box 176, 221 00, Lund, Sweden\\
$^3$Central Glass and Ceramic Research Institute, Kolkata 700032, India
}
\date{\today}
\begin{abstract}
We have investigated magnetoelastic effects in multiferroic YMnO$_3$ below the antiferromagnetic phase transition, $T_N \approx 70$ K, using neutron powder diffraction. The $a$ lattice parameter of the hexagonal unit cell of YMnO$_3$ decreases normally above $T_N$, but decreases anomalously below T$_N$, whereas the $c$ lattice parameter increases with decreasing temperature and then increases anomalously below T$_N$.  The unit cell volume also undergoes an anomalous contraction below $T_N$. By fitting the background thermal expansion for a non-magnetic lattice with the Einstein-Gr\"uneisen equation, we determined the lattice strains $\Delta a$, $\Delta c$ and $\Delta V$ due to the magnetoelastic effects as a function of temperature. We have also determined the temperature variation of the ordered magnetic moment of the Mn ion by fitting the measured Bragg intensities of the nuclear and magnetic reflections with the known crystal and magnetic structure models and have established that the lattice strain due to the magnetoelastic effect in YMnO$_3$ couples with the square of the ordered magnetic  moment or the square of the order parameter of the antiferromagnetic phase transition. 
\end{abstract}
\maketitle
\section{Introduction}
The coupling between spin and lattice degrees of freedom is one of the interesting topics of condensed matter research \cite{callen63,callen65,mayergoyz00,tremolet93,morin90,andreev90,wassermann90,clark80,morosin70,doerr05,lindbaum02}. It has drawn renewed interest in connection with potentially useful electronic materials, such as colossal magnetoresistive and multiferroic materials. One of the most dominant effects of this coupling is the spontaneous exchange striction associated with the magnetic ordering at lower temperatures. If the material does not contain any magnetic ions, the temperature variation of the volume is smooth and shows no anomalies at low temperatures. However, if the material contains magnetic ions that order at low temperature, then at this transition temperature the positions of the atoms may be modified. The most easily probed effects are the modifications of the unit cell parameters and the unit cell volume.  Sometimes the lattice symmetry may get modified following the symmetry of the magnetically ordered state. The symmetry change at the ordering temperatures also causes crystallographic and/or magnetic domains to be formed in single crystal samples. In addition to the lattice strain effects, important modifications in the positional parameters of the atoms and also in the bond distances may take place. However, these modifications are more difficult to detect because the effects are small. To effectively measure this, one needs the temperature variation of the diffraction intensities and, preferably, the diffuse scattering up to high momentum transfer of the scattered radiation. A particularly complex situation arises for magnetic structures with the propagation vector ${\bf k} = 0$. In this case, the magnetic reflections occur at the same position as the nuclear reflections and the refinement of the magnetic and crystal structures become fraught with great difficulties. 

\begin{figure}
\resizebox{0.4\textwidth}{!}{\includegraphics{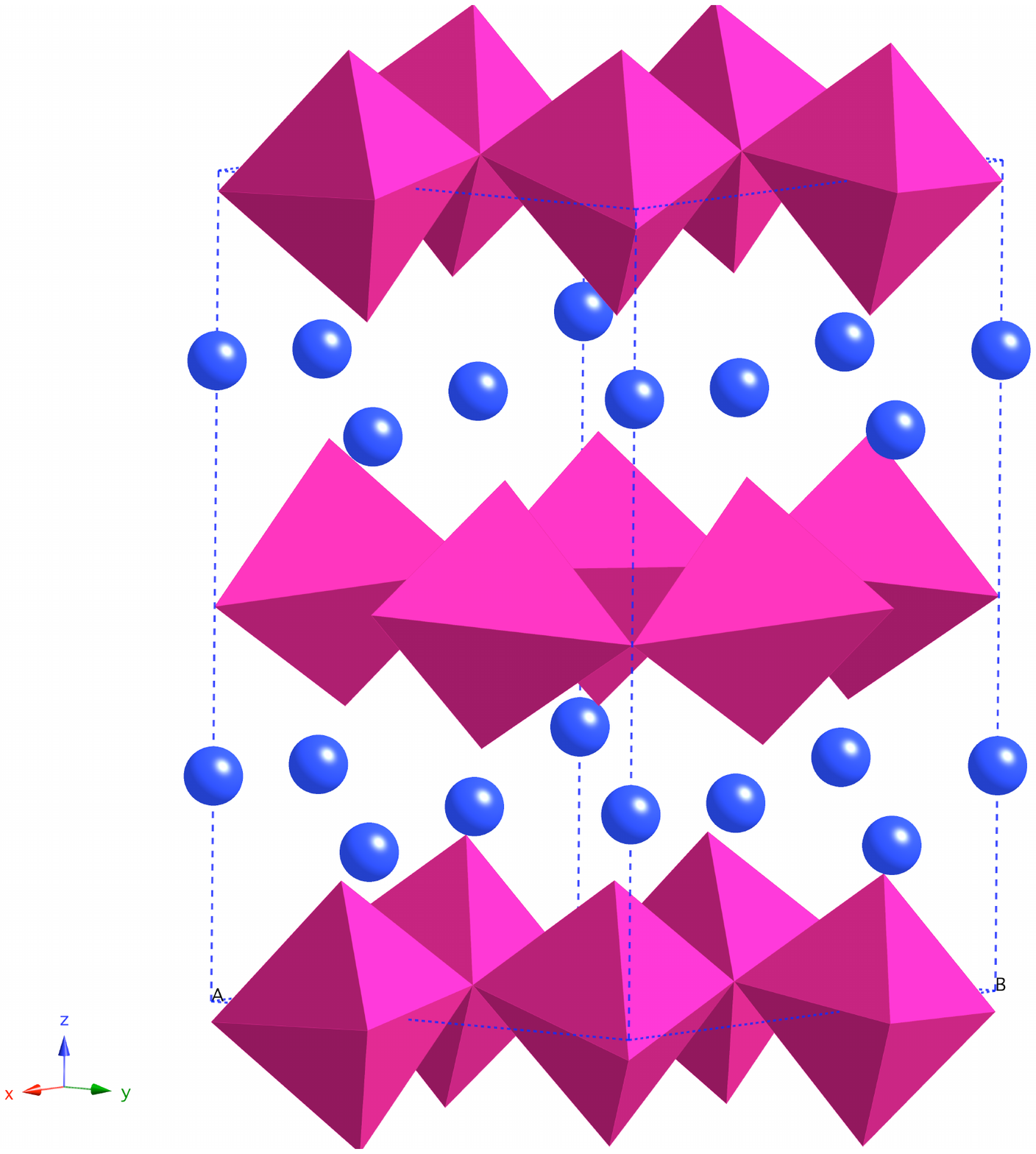}}
\caption {(Color online) Schematic representation of the crystal structure of YMnO$_3$. The red polyhedra represent MnO$_5$ bipyramids and the blue spheres represent the Y ions. }
\label{structure}
\end{figure}

YMnO$_3$ belongs to the family of hexagonal manganites RMnO$_3$ (R = Sc, Y, Er, Ho, Tm, Yb, Lu) that show multiferroic behaviour \cite{huang97}. These hexagonal manganites are paraelectric at high temperatures with the centrosymmetric space group $P6_3/mmc$. Below about 1000 K they undergo a paraelectric-to-ferroelectric transition to the non-centrosymmetric structure with the space group $P6_3cm$. Fig. \ref{structure} shows schematically the crystal structure of YMn$O_3$ in the ferroelectric phase. At lower temperatures, of the order of about 100 K, the magnetic hexagonal manganites order with a non-collinear antiferromagnetic structure with a propagation vector ${\bf k} = 0$. Among these hexagonal manganites YMnO$_3$ and HoMnO$_3$ have been investigated quite intensively \cite{huang97,bertaut63,bertaut67,munoz00,brown06,brown08,vanaken04}. Here we report the powder neutron diffraction investigation of the temperature dependence of the crystal and magnetic structure of YMnO$_3$. We have observed strong magnetoelastic effects or spontaneous magnetostriction below the N\'eel temperature $T_N \approx 70$ K. The lattice strains are found to be proportional to the square of the ordered magnetic moments of the Mn ions. 

In order to determine quantitatively the magnetoelastic effect or the spontaneous magnetostriction, we need to measure the temperature variation of the lattice parameters and the unit cell volume. The high temperature data above the ordering temperature can be fitted to either  a Debye model, using the Gr\"uneisen approximation, \cite{wallace72} or the simpler Einstein model and then extrapolated to lower temperatures to give the background variations in the absence of magnetism. The simpler Einstein-Gr\"uneisen equation serves as an excellent fit function for not only the volume but also for the individual lattice parameters \cite{chatterji11} and we have used it here. 

\begin{figure}
\resizebox{0.5\textwidth}{!}{\includegraphics{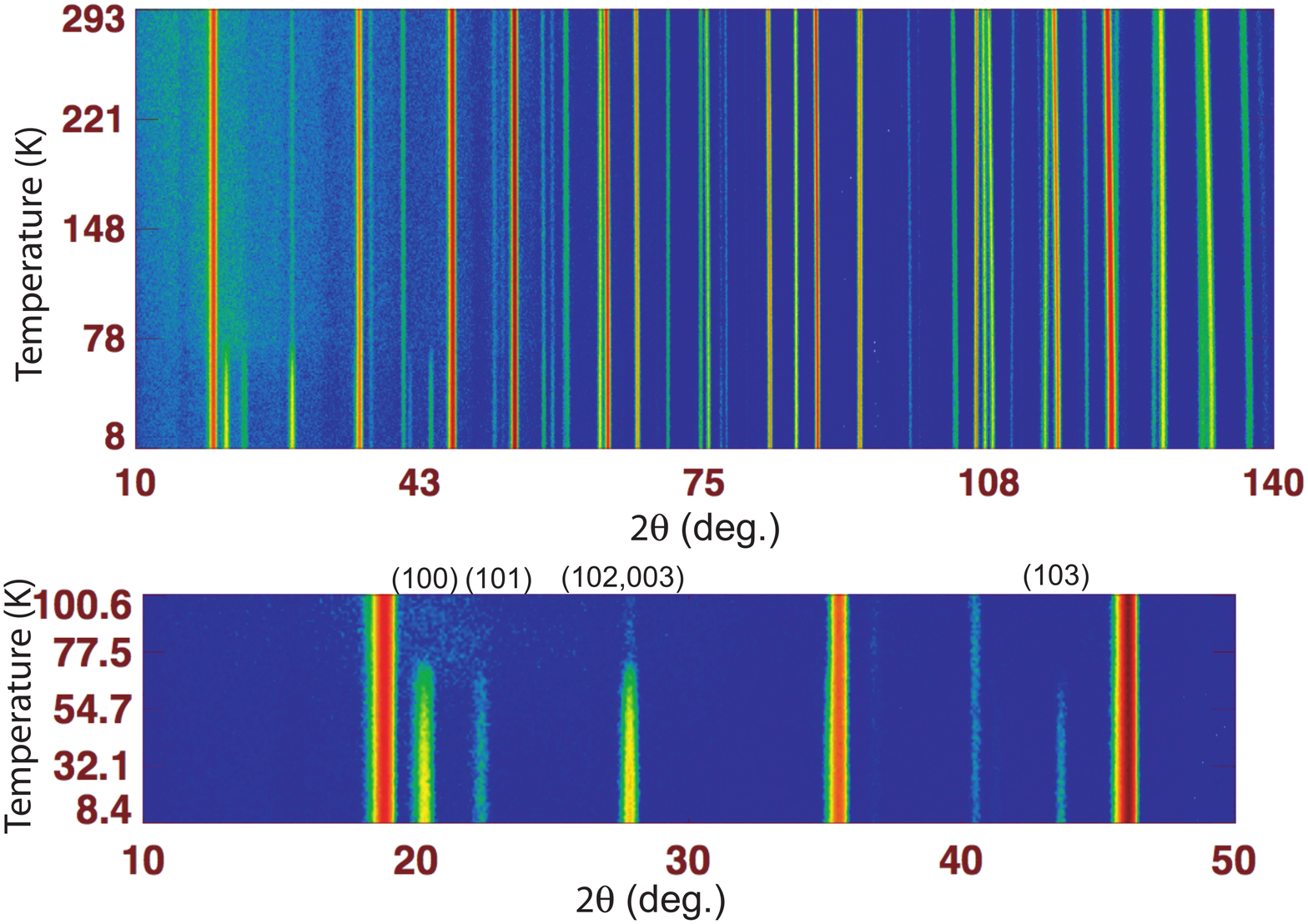}}
\caption {(Color online) (Upper panel)Temperature variation of the diffraction intensities of YMnO$_3$. (Lower panel) The same plot in a limited low  $2\theta$ range showing the temperature variation of the peaks with dominant magnetic contribution.   }
\label{lamp}
\end{figure}

\begin{figure}
\resizebox{0.45\textwidth}{!}{\includegraphics{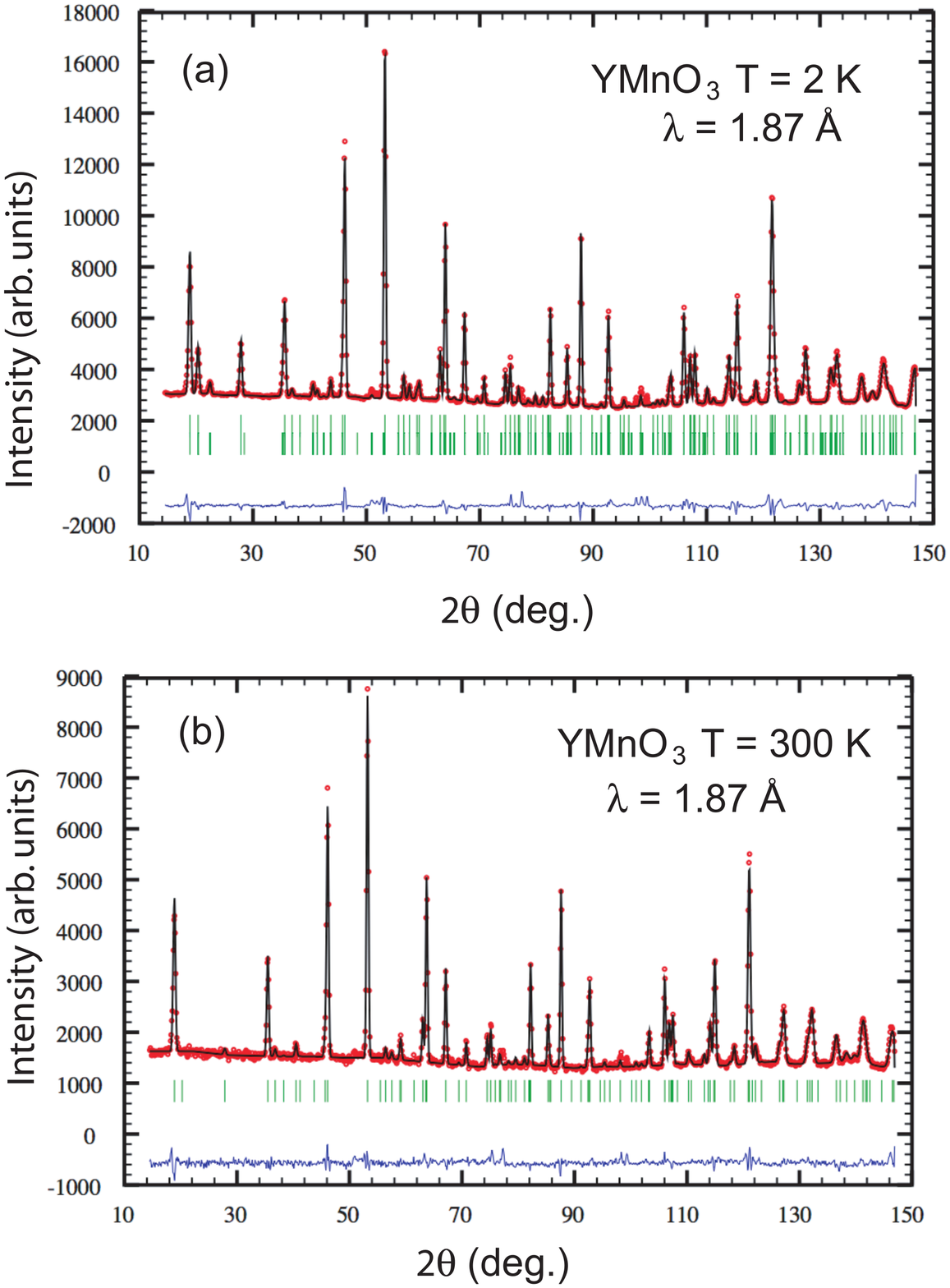}}
\caption {(Color online) Rietveld refinement of the   (a) crystal and magnetic structures of YMnO$_3$ at T = 2 K and (b) the crystal structure of YMnO$_3$ at T = 300 K. The ticks in (a) give the positions of the nuclear and magnetic Bragg peaks whereas those in (b) give the positions of the nuclear peaks only. }
\label{ymnorefinement}
\end{figure}
\begin{figure}
\resizebox{0.5\textwidth}{!}{\includegraphics{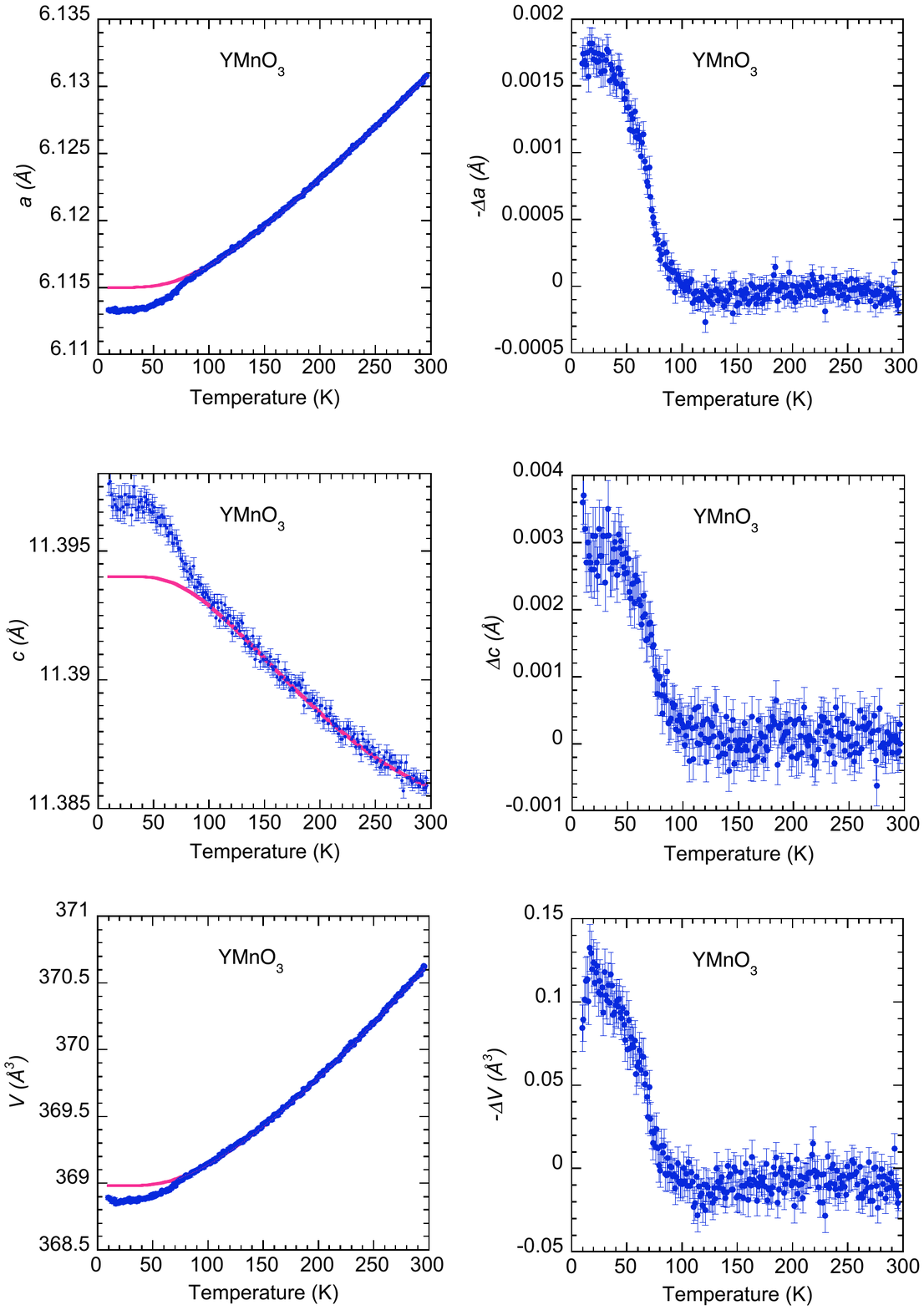}}
\caption {(Color online) Temperature variation of the lattice parameters  $a$, $c$, and the unit cell volume $V$  of YMnO$_3$ plotted on the left panel. The red curves in these figures represent the lattice parameter and the unit cell volume obtained by fitting the high temperature data by the Einstein-Gr\"uneisen equation and extrapolated to low temperature to give the background for the non-magnetic lattice. On the right panel the temperature dependence of the lattice strains $\Delta a$, $\Delta c$ and $\Delta V$ obtained by subtracting the non-magnetic background have been plotted. }
\label{ymnolattice}
\end{figure}
\begin{figure}
\resizebox{0.5\textwidth}{!}{\includegraphics{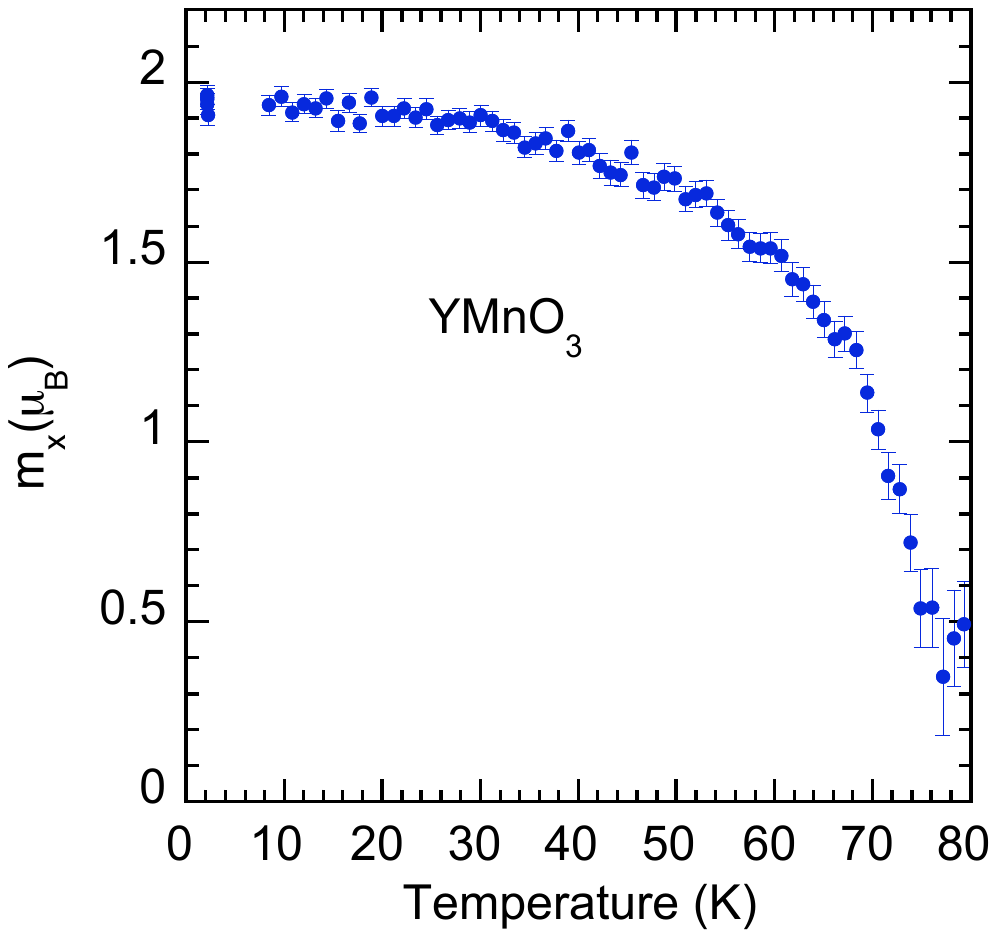}}
\caption {(Color online) Temperature variation of the ordered $x$-component of magnetic moment $m_x$  of the Mn ion obtained by fitting the measured Bragg intensities of the nuclear and magnetic reflections with the known crystal and magnetic structure models.}
\label{moment}
\end{figure}
\begin{figure}
\resizebox{0.5\textwidth}{!}{\includegraphics{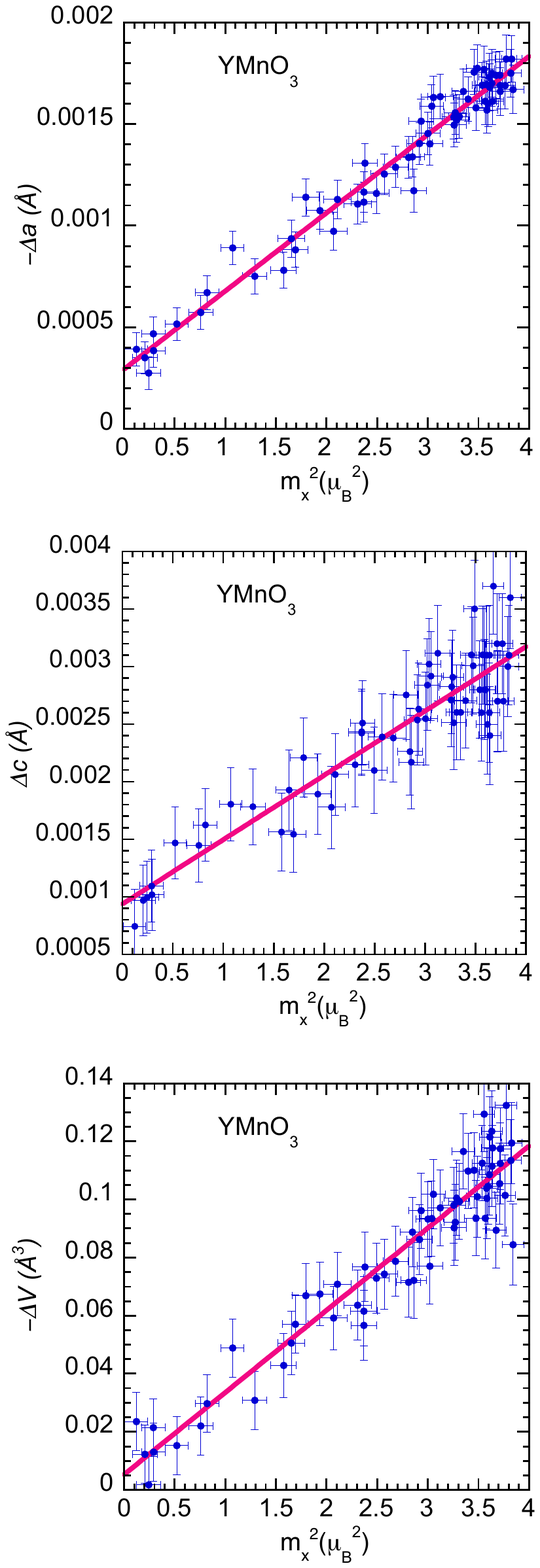}}
\caption {(Color online) Linear plots of the lattice strains $\Delta a$, $\Delta c$ and $\Delta V$ vs. the square of the ordered magnetic moment of the Mn ion in YMnO$_3$.}
\label{linear}
\end{figure}


\section{Experimental Methods}
Neutron diffraction experiments were done on YMnO$_3$ on the high intensity powder diffractometer \cite{hansen08} D20 of the Institute Laue-Langevin in Grenoble. The $115$ reflection from a Ge monochromator at a high take-off angle of $118^{\circ}$ gave a neutron wavelength of 1.868 {\AA}.  Approximately 3 g of powdered YMnO$_3$  was placed inside an $8$ mm diameter vanadium can, which was fixed to the sample stick of a standard $^4$He cryostat. 

\section{Results}
We measured the diffraction intensities from YMnO$_3$ as a temperature in the range $2 - 300$ K. The upper panel of Fig. \ref{lamp} shows the temperature variation of the diffraction intensities of YMnO$_3$ whereas the lower panel shows a similar plot in a limited low  $2\theta$ range showing the temperature variation of the peaks with a dominant magnetic contribution. The magnetic contributions to the intensity are clearly seen in the plot of the lower panel. The reflections like h0l with l = odd disappear completely at $T_N$ as they are forbidden reflections in the space group $P6_3cm$ and have only magnetic contribution and therefore become zero at $T_N$. The other reflections that have both magnetic and nuclear contributions attain a constant value (nuclear) above $T_N$. The Rietveld refinement \cite{rietveld69} of the diffraction data was done using the Fullprof program \cite{rodriguez10}. The refinement results from YMnO$_3$ at T = 2 K  and 300 K are shown in Fig. \ref{ymnorefinement}. At T = 2 K, the intensity data have been fitted with the crystal and the magnetic structure model \cite{brown06} of YMnO$_3$. At T = 300 K, in the paramagnetic state, the refinement was done with the crystal structure model only.

Figure \ref{ymnolattice} shows the temperature variation of the lattice parameters  $a$, $c$, and the unit cell volume $V$  of YMnO$_3$ plotted on the left panel. The lattice parameter $a$ decreases continuously with decreasing temperature but, close to the N\'eel temperature $T_N \approx 70$ K, shows a magnetoelastic or magnetostriction anomaly. The red curve represents the background variation of the $a$ lattice parameter for a non-magnetic solid obtained by fitting the data in the paramagnetic state by the Einstein-Gr\"uneisen equation. By subtracting the background from the data we determined the lattice strain $\Delta a$, plotted on the right panel of Figure \ref{ymnolattice}. The $c$ lattice parameter increases continuously with decreasing temperature down to $T_N$ and then shows an anomalous increase or positive magnetostriction. By subtracting the background variation of the $c$ lattice parameter (red continuous curve) from the data we determined the lattice strain $\Delta c$ plotted on the right panel. Similar plots for the unit cell volume $V$ and and the volume strain $\Delta V$ are shown in Figure \ref{ymnolattice} in the left and the right panels, respectively.  The $c/a$ ratio, which is $1.8572 \pm 0.0001$ at $T = 296$ K, increases linearly with decreasing temperature down to the N\'eel temperature and then attains a constant value $c/a = 1.8643 \pm 0.0001$ at low temperatures. A closer inspection of the temperature variation of the lattice parameters shows that the magnetoelastic effects become actually noticeable at temperatures substantially above T$_N \approx 70$ K. We can tentatively assign a temperature $T^* \approx 100$ K at which the magnetoelastic effect first becomes noticeable. The appearance of the magnetoelastic effect at temperature higher than $T_N$ is presumably due to the short-range spin correlations. The spontaneous linear magnetostriction along the $a$-axis  at T = 0 is $\Delta a/a_0 = -2.944 \times 10^{-4}$ and that along the $c$-axis is $\Delta c/c_0 = 2.633 \times 10^{-4}$. The spontaneous volume magnetostriction $\Delta V/V_0$ is $-3.252 \times 10^{-4}$.  Here $a_0$, $c_0$ and $V_0$ are the extrapolated lattice parameters and the unit cell volumes at $T = 0$ for a fictitious non-magnetic solid obtained by the method described in the previous section. Note that although we determined $\Delta V/V_0 = -3.252 \times 10^{-4}$ by fitting the non-magnetic background and subtracting the experimentally determined unit cell volume, one can also calculate this from the relation $\Delta V/V_0 = 2\Delta a/a_0 +\Delta c /c_0$. The calculated value turns out to be  $\Delta V/V_0 =-3.255 \times 10^{-4}$, very close to the value $-3.252 \times 10^{-4}$ determined semi-independently. This indicates that the method we used for the determination of the non-magnetic background is reliable or at least consistent. One must however point out that the paramagnetic and fictitious non-magnetic background volumes are not really the same but we have assumed this to be the case for simplicity during the present analysis. 

Figure \ref{moment} shows the temperature variation of the $x$-component of the ordered magnetic moment $m_x$  of the Mn ion, obtained by fitting the measured Bragg intensities of the nuclear and magnetic reflections with the known crystal and magnetic structure models. The ordered magnetic moment decreases with increasing temperature and becomes zero at $T_N \approx 75$ K. We did not have enough data points close to $T_N$ and therefore the critical exponent $\beta$ could not be determined.

One expects that the spontaneous magnetostriction is proportional to the square of the ordered magnetic moment. This is at least the case for volume magnetostriction for a single sublattice ferromagnet \cite{andreev95}. However our recent investigations \cite{chatterji10a,chatterji10b} on transition metal difluorides MF$_2$ (M = Mn, Co, Fe, Ni) showed that this is also the case for these simple antiferromagnets with the rutile structure. To check the coupling between the spontaneous magnetostriction and the ordered magnetic moment, or the order parameter of the antiferromagnetic phase transition in YMnO$_3$, we plotted in Fig. \ref{linear} the lattice strains $\Delta a$, $\Delta c$ and $\Delta V$ vs. the square of the $x$-component of the ordered magnetic moment of the Mn ion. The linear plots show that the lattice strains in YMnO$_3$ indeed couple to the square of the magnetic moment or the order parameter. This result also suggests that a second-order spin Hamiltonian is sufficient to describe the magnetic system of YMnO$_3$. It is not necessary to invoke higher order spin Hamiltonian consisting of, for example, four-spin exchange for YMnO$_3$.  

\section{Discussion and conclusions}
Although we were successful in determining the magnetoelastic effects on the lattice parameters, our attempts  to extract the temperature variation of the positional parameters of YMnO$_3$ below $T_N$ did not succeed, presumably because of the ${\bf k} = 0$ magnetic structure of YMnO$_3$. In such structures, magnetic reflections arise below $T_N$ on the same positions as the reflections of structural origin. Well above $T_N$ diffuse magnetic scattering is observed, becoming magnetic Bragg reflections below $T_N$. However, the magnetic scattering in general coincide with nuclear reflections. There are only a few pure magnetic reflections where structural reflections are not allowed due the space group extinctions. The least squares refinement of the crystal and magnetic structures using overlapping nuclear and magnetic intensities at the same reciprocal lattice positions has the problem of separating magnetic and nuclear intensities and therefore lead to correlations between the magnetic and structural parameters. Due to the correlations between structural and magnetic parameters, the positional parameters show abnormal deviations at and near $T_N$. Also these deviations are different if you come from the high  or low temperature sides in a sequential refinement in which one uses the refined parameters of the previous data set as the start parameters of that for the next data set. One way to avoid such unwanted correlations is to use polarized neutron diffraction. The other useful method to disentangle the magnetic and structural parameters is to measure neutron diffraction intensities to a very high momentum transfer $Q = 4\pi\frac{\sin \theta}{\lambda}$. Our present data collected on a reactor based diffractometer at the thermal neutron beam are limited to about $7$ {\AA}$^{-1}$ only. Such high-$Q$ diffraction data can be collected at a spallation neutron source.  One can then refine the crystal structure using only the high-$Q$ data and the magnetic structure using only the low-$Q$ data.  In this way one can hope to get the magnetoelastic effect on the positional parameters (bond distances, symmetric shifts etc.) as well and not just the lattice magnetoelastic effect that we determined in the present paper. Lee et al. \cite{lee08} have obtained magnetoelastic effects of the structural parameters and bond distances of YMnO$_3$, LuMnO$_3$ and Y$_{0.5}$Lu$_{0.5}$MnO$_3$ from their neutron and X-ray diffraction investigations but they have not explained how they have avoided such correlations between the structural and magnetic parameters.  In the absence of such explanation it is difficult to accept the results given by these authors \cite{lee08}. It is extremely important to check the correlation matrix of the refined structural and magnetic parameters, especially in the case of a ${\bf k} = 0$ magnetic structure, because the structural and magnetic intensities appear at the same reciprocal lattice points and least squares refinement process involves dividing the intensities into structural and magnetic parts rather arbitrarily. Only polarized neutron diffraction can separate the magnetic and structural contributions directly and the authors \cite{lee08} did not use such a method. We also checked carefully the supplementary information of the paper in the on-line version. It contains a lot of other information but does not contain the crucial information about the correlation matrix of the least squares refinement.

In conclusion, we have investigated the spontaneous magnetoelastic effects or magnetostrction in multiferroic YMnO$_3$ by neutron powder diffraction. By subtracting the background thermal expansion for a nonmagnetic lattice from the measured thermal expansions we determined the lattice strains $\Delta a$, $\Delta c$ and $\Delta V$ due to the magnetoelastic effects as a function of temperature. We have also determined the temperature variation of the ordered magnetic moment of Mn ion by fitting the measured Bragg intensities of the nuclear and magnetic reflections with the known crystal and magnetic structure models. Our results show that the lattice strain due to the spontaneous magnetostrictions in YMnO$_3$ couples with the square of the ordered magnetic  moment or the square of the order parameter of the antiferromagnetic phase transition. This result also suggests that a second-order spin Hamiltonian is sufficient to describe the magnetic system of YMnO$_3$.

\end{document}